\documentstyle[aps,prb,epsfig]{revtex}

\begin{document}

\twocolumn[\hsize\textwidth\columnwidth\hsize\csname
@twocolumnfalse\endcsname

\title{Nonlinear interaction of charged particles with a free electron
gas beyond the random-phase approximation}
\author{T. del R\'\i o Gaztelurrutia$^{1}$ and J. M. Pitarke$^{2,3}$}
\address{$^1$ Fisika Aplikatua Saila, Industri eta Telekomunikazio Ingeniarien
Goi Eskola Teknikoa,\\
Urkijo Zumarkalea z/g, S-48013 Bilbo, Spain\\ 
$^2$ Materia Kondentsatuaren Fisika Saila, Zientzi Fakultatea, 
Euskal Herriko Unibertsitatea,\\ 644 Posta kutxatila, 48080 Bilbo, Basque 
Country, Spain\\
$^3$ Donostia International Physics Center (DIPC) and Centro Mixto
CSIC-UPV/EHU,\\ Donostia, Basque Country, Spain}

\date{\today}

\maketitle

\begin{abstract}
A nonlinear description of the interaction of charged particles
penetrating a solid has become of basic importance in the interpretation of
a variety of physical phenomena. Here we develop a many-body theoretical
approach to the quadratic decay rate, energy loss, and wake potential of
charged particles moving in an interacting free electron gas. Explicit
expressions for these quantities are obtained either within the random-phase
approximation (RPA) or with full inclusion of short-range exchange and
correlation effects. The $Z_1^3$ correction to the energy loss of ions is
evaluated beyond RPA, in the limit of low velocities.  
\end{abstract}

]

When charged particles pass through a solid, energy can
be lost to the medium through various types of elastic and inelastic collision
processes.\cite{Echenique} While at relativistic velocities radiative losses may
become important, for moving charged particles in the non-relativistic regime the
energy loss is primarily due to electron-electron (e-e) interactions giving rise
to the generation of electron-hole pairs, collective excitations such as
plasmons, and inner-shell excitations and ionizations. Energy losses due to
nuclear recoil are negligible, unless the projectile velocity is
very small compared to the mean speed of electrons in the solid.\cite{Komarov}

The inelastic decay rate of charged particles in a degenerate interacting
free electron gas (FEG) has been calculated for many years in 
the first-Born approximation or, equivalently, within linear-response
theory. This is a good approximation when the velocity of the
projectile is much greater than the average velocity of target electrons.
However, in the case of projectiles moving with smaller velocities,
nonlinearities have been shown to play a key role in the interpretation of a
variety of experiments. Energy-loss measurements have revealed differences, not
present within linear-response theory, between the energy loss of protons and
antiprotons.\cite{antiproton,Moller} Moreover, experimentally
observed coherent double-plasmon excitations\cite{doublep,doublepp} cannot be
described within linear-response theory, and nonlinearities may also play an
important role in the electronic wake generated by moving ions in a
FEG.\cite{Vager}

Pioneering nonlinear calculations of the electronic energy loss of low-energy
ions in an electron gas were performed by Echenique {\it et
al\,}.\cite{Echenique1} These authors computed the scattering cross section for a
statically screened potential, which was determined self-consistently using
density-functional theory (DFT).\cite{DFT} These static-screening
calculations have recently been extended to velocities approaching the Fermi
velocity.\cite{Salin} Second-order perturbative calculations, which do not have
the limitation of being restricted to low projectile velocities, have been
reported by different authors with use of the random-phase approximation (RPA)
and by treating the moving charged particle as a prescribed source of energy and
momentum.\cite{Sung,Zaremba,Esbensen,Pitarke1,Pitarke2}

In this paper, we report a many-body theoretical approach to the
quadratic decay rate, energy loss, and wake potential of charged particles
moving in an interacting FEG. The decay rate is derived from the knowledge of
the projectile self-energy. The energy loss and wake potential are obtained
within quadratic response theory. While the first-order contribution to the
energy loss may also be obtained from the imaginary part of the projectile
self-energy by simply inserting the energy transfer inside the integrand of this
quantity, our results indicate that this procedure cannot be generalized to the
description of the second-order energy loss. Unless otherwise is stated, we
use atomic units throughout, i.e., $e^2=\hbar=m_e=1$.

We consider the interaction of a moving probe particle of charge $Z_1$ and mass
$M$ with a FEG of density $n$. The probe particle is assumed to be
distinguishable from the electrons in the Fermi gas, which is described by
an isotropic homogeneous assembly of interacting electrons immersed in a uniform
background of positive charge and volume $V$. Many-body theory shows that the
probability for the probe particle to occupy a given excited state of
four-momentum $p$ decays exponentially in time with the decay
constant\cite{Fetter}
\begin{equation}\label{tau0}
\tau^{-1}=-2\,{\rm Im}\Sigma_p,
\end{equation}
where $\Sigma_p$ represents the particle self-energy.

It is well known that the self-energy cannot be computed by
simply evaluating the lowest-order Feynman diagrams, because of
severe infrared divergences due to the long-range Coulomb interaction. Instead,
one needs to resum electron-loop corrections and expand in
terms of the dynamically screened interaction. Up to third order in
$Z_1$, the self-energy of the probe particle can be represented
diagrammatically as shown in Fig. 1. The sum of the first two diagrams represents the so-called
$GW$ approximation, and the third diagram accounts for $Z_1^3$ corrections. One
finds
\begin{eqnarray}\label{auto1}
&&\Sigma_p=i\,Z_1^2\int{dq^4\over(2\pi)^4}\,v_{\bf q}\,D_{p-q}
\left[(1+\chi_q\,v_{\bf q})\right.\cr\cr
&&\left.-2\,i\,Z_1\int{d^4q_1\over(2\pi)^4}\,
D_{p-q_1}\,D_{p-q+q_1}\,Y_{q,-q_1}\,v_{{\bf q}_1}v_{{\bf q}-{\bf q}_1}\right],
\end{eqnarray}
where $v_{\bf q}$ is the Fourier transform of the bare Coulomb
potential, $D_p$ is the probe-particle propagator
\begin{equation}
D_p={1\over p^0-\omega_{\bf p}-\Sigma_p+i\eta},
\end{equation}
$\omega_{\bf p}={\bf p}^2/(2M)$ is the noninteracting energy, and $\eta$ is a
positive infinitesimal.
$\chi_q$ and $Y_{q_1,q_2}$ represent {\it exact} time-ordered density
correlation functions of the interacting FEG,
\begin{equation}\label{eq17}
\chi_q={1\over V}\sum_n\left|(\rho_{\bf
q})_{n0}\right|^2\left[{1\over q^0+\omega_{0n}+{\rm
i}\eta}-{1\over q^0-\omega_{0n}-i\,\eta}\right]
\end{equation}
and
\begin{eqnarray}\label{eq2p}
Y_{q_1,q_2}=-{1\over 2V}
\sum_{n,l}&&\left[\,{(\rho_{{\bf q}_1})_{0n}(\rho_{{\bf
q}_3})_{nl}(\rho_{{\bf q}_2})_{l0}\over (q^0_1+\omega_{0n}+{\rm
i}\eta)(q^0_2+\omega_{l0}-i\,\eta)}\right.\cr\cr
&&\left.+{(\rho_{{\bf
q}_2})_{0n}(\rho_{{\bf q}_1})_{nl}(\rho_{{\bf q}_3})_{l0}\over
(q^0_2+\omega_{0n}+i\,\eta)(q^0_3+\omega_{l0}-{\rm
i}\eta)}\right.\cr\cr
&&\left.+{(\rho_{{\bf q}_3})_{0n}(\rho_{{\bf
q}_2})_{nl}(\rho_{{\bf q}_1})_{l0}\over (q^0_3+\omega_{0n}+{\rm
i}\eta)(q^0_1+\omega_{l0}-i\,\eta)}\right.\cr\cr
&&\left.+(q_2\rightarrow q_3)\right],
\end{eqnarray}
$\left(\rho_{\bf q}\right)_{nl}$ being the matrix element of the Fourier
transform of the electron-density operator, taken between exact many-electron
states of energy $E_n$ and $E_l$, $\omega_{nl}=E_n-E_l$, and $q_3=-(q_1+q_2)$.

If the probe particle is an ion ($M>>1$), the propagator $D_p$ and
the energy $p^0$ entering Eq. (\ref{auto1}) can be safely approximated by the
noninteracting propagator $D_p^0$ and energy $\omega_{\bf p}$. Recoil can also
be neglected, and the introduction of Eq. (\ref{auto1}) into Eq. (\ref{tau0})
then yields, after some work of rearrangement, the following expression:
\begin{eqnarray}\label{tau2}
&&\tau^{-1}=4\pi\,Z_1^2\int{dq^4\over(2\pi)^4}\,v_{\bf q}\,\delta(q^0-{\bf
q}\cdot{\bf v})\,\Theta(q^0)\cr\cr
&&\times\left[-{\rm Im}K_q+{4\over 3}\,\pi\,Z_1\int{d^4q_1\over(2\pi)^4}\,{\rm
Im}Y_{q,-q_1}\,v_{{\bf q}_1}v_{{\bf q}-{\bf q}_1}\right.\cr\cr
&&\left.\quad\times\delta(q_1^0-{\bf q}_1\cdot{\bf v})\right],
\end{eqnarray}
where ${\bf v}$ is the particle velocity, $\Theta(x)$ represents the Heaviside
step function, and $K_q$ is the so-called test$\_$charge-test$\_$charge inverse
dielectric function:
\begin{equation}\label{inverse}
K_q=1+\chi_q\,v_{\bf q}.
\end{equation}
The decay rate of Eq. (\ref{tau2}) has not been reported before.

In the RPA, density correlation functions are obtained
by summing over all ring-like diagrams,
\begin{equation}
\chi^{\rm RPA}_q=\chi_q^0+\chi_q^0\,v_{\bf q}\,\chi^{RPA}_q
\end{equation}
and
\begin{equation}\label{RPA2}
Y_{q_1,q_2}^{RPA}=
K_{q_1}^{RPA}\,Y_{q_1,q_2}^0\,K_{-q_2}^{RPA}\,K_{-q_3}^{RPA},
\end{equation}
where $\chi_q^0$ and $Y_{q_1,q_2}^0$ represent noninteracting density
correlation functions. Improvements on the RPA are typically carried out by
introducing an effective e-e interaction\cite{Singwi}
\begin{equation}\label{eqHub}
\tilde v_q=v_{\bf q}\left(1-G_q\right),
\end{equation}
where $G_q$ is the so-called local-field factor accounting for short-range
exchange and correlation (xc) effects not present in the RPA. Accordingly, the
density correlation functions
$\chi_q$ and
$Y_{q_1,q_2}$ are found to be of the RPA form, but with all e-e bare-Coulomb
interactions
$v_{\bf q}$ replaced by $\tilde v_q$, i.e.,
\begin{equation}\label{chi2}
\chi_q=\chi_q^0+\chi_q^0\,\tilde v_q\,\chi_q
\end{equation}
and
\begin{equation}\label{Y2}
Y_{q_1,q_2}=
\tilde K_{q_1}\,Y_{q_1,q_2}^0\,\tilde K_{-q_2}\,\tilde K_{-q_3},
\end{equation}
where $\tilde K_q$ is the test$\_$charge-electron inverse dielectric
function\cite{Kleinman,HL}
\begin{equation}\label{inversep}
\tilde K_q=1+\chi_q\,\tilde v_q.
\end{equation}
In the RPA, $K_q$ and $\tilde K_q$ coincide.

The potential induced in a uniform FEG by the presence of the recoiless
probe particle may be obtained with the use of time-dependent perturbation
theory. Keeping terms of first and second order in the external perturbation,
one finds
\begin{eqnarray}\label{vind}
&&V^{ind}({\bf r},t)=2\pi\,Z_1\int{d^4 q\over(2\pi)^4}\,
{\rm e}^{i\left({\bf q}\cdot{\bf r}-q^0\,t\right)}\,v_{\bf
q}\cr\cr
&&\times\delta\left(q^0-{\bf q}\cdot{\bf v}\right)
\left[(K_q^R-1)-2\pi\,Z_1\int{d^4
q_1\over(2\pi)^4}\,Y_{q,-q_1}^R\right.\cr\cr
&&\left.\quad\quad\quad\times v_{{\bf q}_1}\,v_{{\bf q}-{\bf
q}_1}\,\delta\left(q_1^0-{\bf q}_1\cdot{\bf v}\right)\right],
\end{eqnarray} 
where we have introduced the retarded counterparts of the time-ordered
functions $K_q$ and $Y_{q,-q_1}$ entering Eq. (\ref{tau2}).

The average energy lost per unit length traveled by the probe particle
is obtained as the retarding force due to the potential of Eq. (\ref{vind})
induced in the vicinity of the projectile itself. One easily finds
\begin{eqnarray}\label{eloss}
-{dE\over dx}=&&4\pi\,Z_1^2\int{d^4
q\over(2\pi)^4}\,q^0\,v_{\bf q}\,\delta\left(q^0-{\bf q}\cdot{\bf
v}\right)\,\Theta(q^0)\cr\cr
&&\times\left[-{\rm Im}K_q^R+2\pi\,Z_1\int{d^4 q_1\over(2\pi)^4}\,{\rm
Im}Y_{q,-q_1}^R\right.\cr\cr
&&\left.\quad\times v_{{\bf q}_1}\,v_{{\bf q}-{\bf
q}_1}\,\delta\left(q_1^0-{\bf q}_1\cdot{\bf v}\right)\right].
\end{eqnarray}

Both Eq. (\ref{tau2}) for the decay rate and Eq. (\ref{eloss}) for the energy
loss can be derived, within many-body perturbation theory, from the knowledge of
the probability for the probe particle to transfer four-momentum $q$ to the
FEG.\cite{Teresa} Nevertheless, this probability cannot be identified with the
integrand of Eq. (\ref{tau2}), and therefore the energy loss of Eq.
(\ref{eloss}) cannot be obtained by simply inserting the energy transfer inside
the integrand of Eq. (\ref{tau2}).  

In the framework of time-dependent density-functional theory
(TDDFT),\cite{Peter} quadratic response theory yields
\begin{equation}\label{chil}
\chi_{q}^R=\chi^{R,0}_{q}+\chi^{R,0}_{q}\,(v_{\bf q}+f_q^{xc})\,\chi_{q}^R,
\end{equation}
\begin{equation}\label{chiq}
Y_{q_1,q_2}^R=\tilde K_{q_1}^R\,Y_{q_1,q_2}^0\,\tilde
K_{-q_2}^R\,\tilde K_{-q_3}^R,
\end{equation}
\begin{equation}\label{inversep}
K_q^R=1+\chi_q^R\,v_{\bf q},
\end{equation}
and
\begin{equation}\label{inversetilde}
\tilde K_q^R=1+\chi_q^R\,(v_{\bf q}+f_q^{xc}),
\end{equation}
where $\chi_q^{R,0}$ and $Y_{q_1,q_2}^{R,0}$ represent retarded noninteracting
density correlation functions, and $f_q^{xc}$ denotes the Fourier transform of 
\begin{equation}
f^{xc}(x,x')={\delta V^{xc}([n],x)\over\delta n(x')},
\end{equation}
$V^{xc}([n],x)$ being the exact time-dependent xc
potential of TDDFT.\cite{Peter}. Within RPA $f_q^{xc}=0$, and introduction of
Eqs. (\ref{chiq}) and (\ref{inversep}) into Eqs. (\ref{vind}) and
(\ref{eloss}) then yields the results derived
in Refs.\onlinecite{Sung,Zaremba,Esbensen,Pitarke1,Pitarke2}.

At this point, we present an application of our formalism, namely the
low-velocity limit of the energy loss with inclusion of short-range xc
effects. For low projectile velocities ($v\to 0$), only the static
($\omega\to 0$) $\chi_q^{R,0}$, $Y_{q_1,q_2}^{R,0}$, and $f_q^{xc}$
enter in the evaluation of the energy loss of Eq. (\ref{eloss}), which is then
easily found to be proportional to the projectile velocity. In particular, in the
so-called local-density approximation (LDA), which is rigorous in the
long-wavelength limit ($q\to 0$), one finds
\begin{equation}\label{eq32}
f_q^{xc}={4\pi\over q_F^2}\left[{1\over 4}-{4\pi\over
q_F^2}{d^2E_{c}\over dn^2}\right],
\end{equation}
$E_c(n)$ being the correlation energy of a uniform electron gas of density $n$.

Diffusion Monte Carlo calculations of $\chi_q^0$ have
shown that the LDA static xc kernel of Eq. (\ref{eq32}) reproduces
correctly the static response  for all $q\le 2\,q_F$.\cite{Alder} We have
calculated static density correlation functions from Eqs. (\ref{chil}) and
(\ref{chiq}), with use of the LDA static xc kernel of Eq. (\ref{eq32}) and the
Perdew-Zunger\cite{Perdew} parametrization of the quantum Monte Carlo
correlation energy $E_c(n)$ of Ceperley and Alder.\cite{CA}

In Fig. 2 we show linear ($\propto Z_1^2$) and quadratic ($\propto Z_1^3$)
contributions to the low-velocity energy loss of Eq. (\ref{eloss})
divided by the velocity of the projectile, as a function
of the electron-density parameter $r_s$.\cite{rs} Comparison between our {\it
new} results (solid and short-dashed lines) and those obtained within RPA
(long-dashed and dotted lines) indicates that xc effects become increasingly
important as the electron density decreases, the impact of these effects being
more pronounced for the
$Z_1^3$ than for the $Z_1^2$ contribution to the energy loss. As a result, the
importance of the quadratic contribution increases when xc effects are included,
and for $r_s\sim 2.5$ it is equal in magnitude to the linear contribution [within
RPA, linear and quadratic contributions are equal  for $r_s\sim 5$].

The crosses and rhombs in Fig. 2 represent the full nonlinear contribution
to the energy loss [difference between the total energy loss and the
linear contribution], multiplied by a factor of $-1$, as obtained from DFT
calculations for antiprotons to all orders in $Z_1$,\cite{Nagy} with xc
effects excluded (crosses) and with LDA xc effects included (rhombs). Both full
nonlinear calculations are found to agree nicely with our quadratic-response
calculations in the high-density limit ($r_s\to 0$).
As the electron density decreases, quadratic-response calculations overestimate
the energy loss of antiprotons, especially when xc effects are included,
showing that xc and nonlinear (beyond $Z_1^3$) effects tend to compensate.

Finally, we note that linear and quadratic contributions to the energy loss
can be extracted from the small-$Z_1$ behaviour of full nonlinear DFT
calculations, which have been reported in the low-velocity
limit.\cite{Echenique1} When
$(-dE/dx)/Z_1^2$ is plotted as a function of
$Z_1$, the result has a linear $Z_1$ dependence at small $Z_1$. The
intercept and the slope of this curve at $Z_1=0$ give linear and quadratic
contributions to the full nonlinear energy loss, as shown in
Ref.\onlinecite{Zaremba} by simply ignoring xc effects. These authors
repeated their full nonlinear DFT calculations with LDA xc included, but were
unable to compare them with quadratic-response calculations that included xc
effects at the same level of approximation. The results of these
quadratic-response calculations are now shown in Fig. 2. It can be seen that the
ratio between xc-included (short-dashed line) and xc-excluded (dotted line) 
quadratic contributions to the energy loss is in excellent agreement with the
ratios reported in Ref.\onlinecite{Zaremba} [see Fig. 7 of this reference] as
derived from the $Z_1$-dependence of full nonlinear DFT calculations.

In conclusion, we have developed a many-body theoretical approach to
the quadratic decay rate, energy loss, and wake potential of charged particles
moving in an electrons gas, with full inclusion of short-range xc effects. We
have shown that in the limit of high electron densities and low projectile
charges our calculated quadratic energy loss, which can be extended to the
case of larger velocities, reproduces DFT calculations for antiprotons, as long
as exchange and correlation are treated at the same level of accuracy.

\acknowledgments

The authors acknowledge partial support by the University of the Basque
Country, the Basque Hezkuntza, Unibertsitate eta Ikerketa Saila, and the
Spanish Ministerio de Educaci\'on y Cultura.

\begin{figure}
\center{
 \epsfig{file=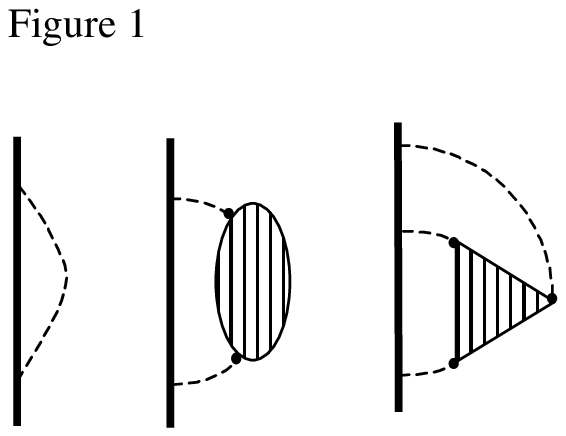}}
\caption{The probe-particle self-energy, up to third order in $Z_1$. Thick
solid lines represent the exact probe-particle propagator, $i\,D_p$. Dashed
lines represent the bare Coulomb interaction, $-i\,v_{\bf q}$. Two- and three-point loops represent time-ordered
density correlation functions, $i\,\chi_q$ and $-2\,Y_{q_1,q_2}$, respectively.}
\end{figure}

\begin{figure}

\caption{$Z_1^2$ and $Z_1^3$ contributions to the low-velocity
electronic stopping power of Eq. (\ref{eloss}) divided by the velocity of the
projectile, as a function of $r_s$, for $Z_1=1$. Our full calculations, as
obtained with inclusion of LDA exchange and correlation, are
represented by solid ($Z_1^2$ contribution) and short-dashed ($Z_1^3$
contribution) lines. Long-dashed ($Z_1^2$) and dotted ($Z_1^3$) lines represent
the corresponding results obtained within RPA ($f_q^{xc}=0$). The difference
[multiplied by a factor of
$-1$] between the full nonlinear stopping power for antiprotons, as reported in
Ref.\protect\onlinecite{Nagy}\protect, and the linear contribution is represented
by crosses (xc effects excluded) and rhombs (LDA xc included).}
\end{figure}

\epsfig{file=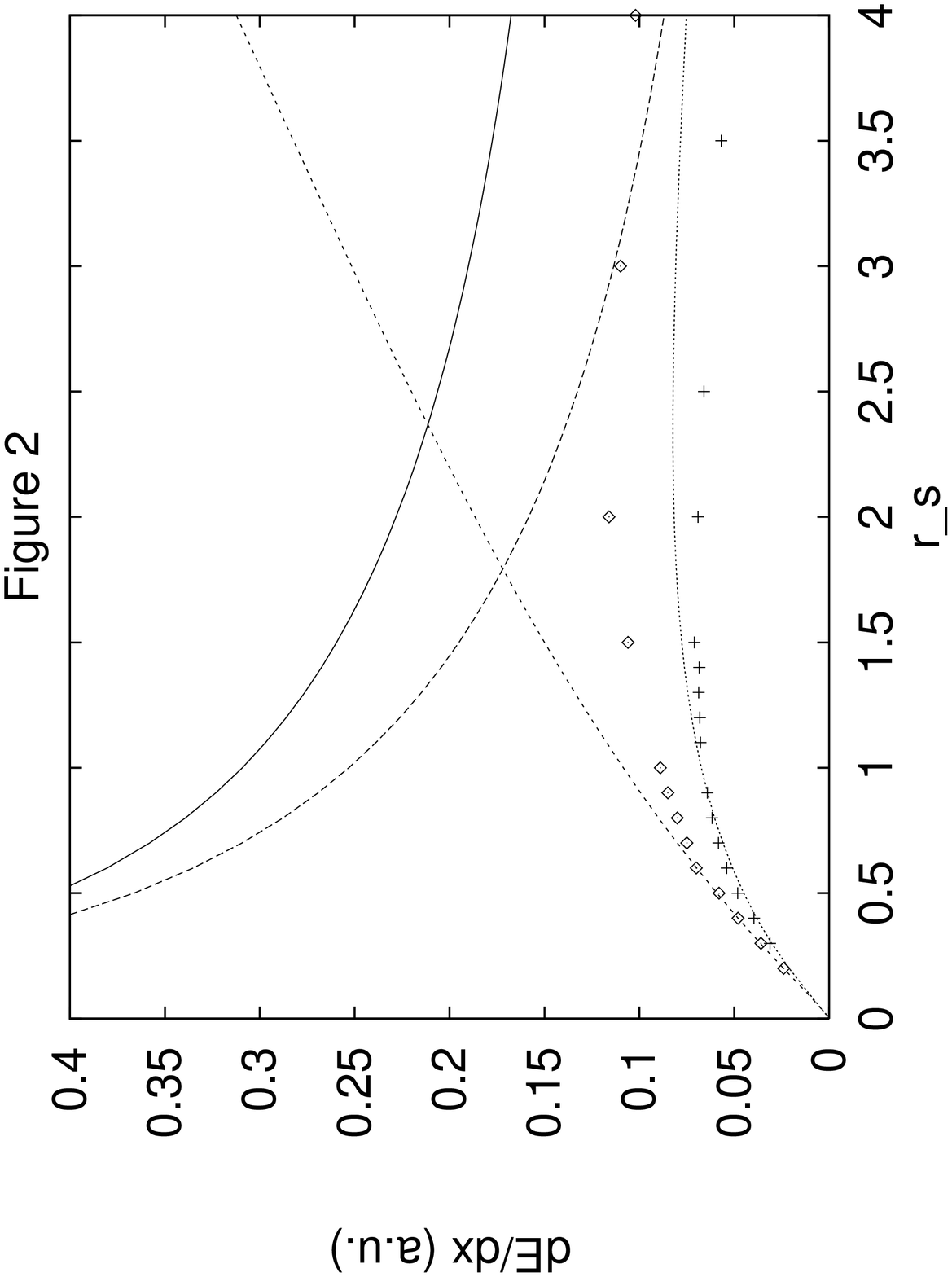}

\end{document}